\documentstyle[12pt]{article}
\begin{document}
\title{GRAVITATIONAL INSTABILITY FOR A MULTIFLUID MEDIUM IN AN EXPANDING 
UNIVERSE}
\author{D. Fargion \\
Phys. Dept. University of Rome ``La Sapienza''.INFN}
\date{September 11 1983}
\maketitle
\begin{abstract}
We consider the gravitational clustering of a multicomponent fluid in an 
expanding Newtonian universe , taking into account the mutual gravitational 
interactions of the medium. We obteined a set of exact solutions for two fluid 
components and for several fluid components : primordial hydrigen and helium 
or cosmological massive neutrinos of different masses and/or velocity 
distribution may be described by these solutions. The ratio between the two 
fluid perturbations , even if originally equal to unity , changes 
monotonically with time; this phenomenon leads to different epochs for 
nonlinear clustering that may be reflected into different epochs of galaxy 
formation. 
\end{abstract}
\section*{Introduction}
Following the Newtonian approximation suggested by BONNOR \cite{bonnor} and 
L.P. Grishchuk and Zel'dovich \cite{gris} for the gravitational instability in 
a multicomponent hydrodynamic medium , let us consider , upon two initially 
homogeneous and stationary fluid densities $\rho_{1}$ , $\rho_{2}$ and 
pressures $p_{1}$ , $p_{2}$ , two small perturbations represented by plane 
waves with wave vector $\vec{K}$ :
\begin{equation}
\frac{\delta \rho_{1}}{\rho_{1}} = \delta_{1}(t) e^{i \vec{K} 
\cdot \, \vec{r}}
\;\;\; , \;\;\; \frac{\delta \rho_{2}}{\rho_{2}} = \delta_{2}(t) 
e^{i \vec{K} \cdot \, \vec{r}} \;.
\end{equation}
the first linear approximation of adiabatic fluctuations with velocities of 
sound
\begin{equation}
v^{2}_{1s} = \frac{d p_{1}}{d \rho_{1}} \;\;\;\; , \;\;\; 
v^{2}_{2s} = \frac{d p_{2}}{d \rho_{2}}
\end{equation}
has been already studied exhaustively by Grishchuk and Zel'dovich for a 
stationary universe.\\
Here we take into account the expansion of the Universe using the non-
relativistic Newtonian treatment , considering the following system of coupled 
equations :
\begin{equation}
   \left[  
   \begin{array}{l}
\ddot{\delta_{1}} + 2 \frac{\dot{R}}{R} \dot{\delta_{1}} 
   + \left( v^{2}_{1s} K^{2} - 4 \pi G \rho_{1} \right) \delta_{1} 
   - 4 \pi G \rho_{2} \delta_{2} = 0 \\
   \\
\ddot{\delta_{2}} + 2 \frac{\dot{R}}{R} \dot{\delta_{2}}
   + \left( v^{2}_{2s} K^{2} - 4 \pi G \rho_{2} \right) \delta_{2}
   - 4 \pi G \rho_{1} \delta_{1} = 0   
   \end{array}
   \right.
\end{equation}
where the dot stands for the time derivative and $R$ is the cosmological scale 
factor ; $K^{2} = \left| \vec{K} \right|^{2}$.\\
System (3) reduces to the same one considered by Grishchuk and Zel'dovich 
\cite{gris} for the stationary solution $\left( \dot{R} = 0 \right)$ and it 
also contains the system studied by Wasserman \cite{Was} , in the limit 
$ K^{2} = 0$.\\
Let us summarize the previous results concerning the stationary universe :
\begin{itemize}
\item i) There is no more than one unstable mode of exponential form 
$ \delta \propto e^{w t} , w > 0 \;$ occurring for wave vectors $K$ which 
satisfy 
\begin{equation}
K^{2} < K^{2}_{j} \equiv \frac{4\pi G \rho_{1}}{v^{2}_{1s}} + 
                         \frac{4\pi G \rho_{2}}{v^{2}_{2s}} \;\;\; .
\end{equation}
\item ii) The exponent $w$ should satisfy the relation
\[
\omega^4 + w^2 \left[ K^2 \left( v^2_{1s} + v^2_{2s} \right) - 4 \pi G 
       \left( \rho_{1} + \rho_{2} \right) \right] + 
\]
\begin{equation}
\;\;\;\;\;\;\;\;\;\;\;\;\;\;\;\; + K^2 \left[ K^2 v^2_{1s} v^2_{2s} - 4 \pi G 
       \left( \rho_{1} v^2_{2s} + \rho_{2} v^2_{1s} \right) \right] = 0 \; .
\end{equation}
\item iii) The basic laws governing the gravitational instability in a 
two-component medium hold true for a multicomponent medium as well.
\end{itemize}
The following results for an expanding universe show analogous behaviours.
We note that eq (3) may not be applied to much earlier epochs than the 
recombination , when cosmological expantion was radiation dominated. However , 
at that time the radiation drag cuts off very effectively the non-relativistic 
instability \cite{Pee}.\\
Therefore , we restrict our attention to the system of equations (3) for the 
cosmological epochs dominated by matter.\\
As present cosmological curvature is near to unity , at earlier times one may 
approximate with good accurancy the cosmological evolution to a flat Friedmann 
universe ; therefore ,
\begin{equation}
\left[
\begin{array}{lll}
H = \frac{\dot{R}}{R} = \frac{2}{3} \, t^{-1} & 
4\pi G \rho_{i} = \frac{2}{3} \,\frac{\Omega_{i}}{\Omega} \, t^{-2} \; , & \\
\\
\Omega_{i} = \frac{\rho_{i}}{\rho_{c}} & \Omega = 
   \frac{\rho_{1} + \rho_{2}}{\rho_{c}} \leq 1 & \rho_{c} = 
   \frac{3 H^2}{8 \pi G} \;.
\end{array}
\right.
\end{equation}
In particular , $\Omega = 1 $ if one disregards all the other 
contributions to the matter density out of the ones for fluids 1 , 2.\\
Given a specific-heat ratio $\gamma_{i}$ for each fluid component , according 
to Weinberg \cite{wein} and eq. (6) , the speeds of sound may be written as 
follows:
\begin{equation}
v_{is} = \sqrt{\frac{\gamma_{i} p_{i}}{\rho_{i}}\,} \, \propto \, 
         \rho^{\frac{1}{2} \left( \gamma_{i} - 1 \right)} \, \propto \,
         t^{1 - \gamma_{i}} \;\;\; , i = 1 , 2 \;\;. 
\end{equation}
In particular , we choose $v_{is} = v_{i} \, t^{1 - \gamma_{i}} \,$ , where 
$v_{i}$ are constants.\\
The wave vector $\vec{K}$ decreases like the inverse of the cosmological 
radius $R$ and from eq. (6)  we may write $K^2 = k^2 \, t^{-\frac{4}{3}} \,$ , 
where $k$ is a constant.\\
Consequently , system (3) becomes
\begin{equation}
\left[
\begin{array}{l}
\ddot{\delta_{1}} + \frac{4}{3} \, \frac{\dot{\delta_{1}}}{t} + 
     \left[ \frac{k^2 v^2_{1}}{t^{\left(2 \gamma_{1} - \frac{2}{3}\right)}} 
     - \frac{2}{3} \frac{\Omega_{1}}{\Omega t^2} \right] \delta_{1} - 
     \frac{2}{3} \, \frac{\Omega_{2}}{\Omega} \frac{\delta_{2}}{t^2} = 0 \\
\\
\ddot{\delta_{2}} + \frac{4}{3} \, \frac{\dot{\delta_{2}}}{t} +
     \left[ \frac{k^2 v^2_{2}}{t^{\left(2 \gamma_{2} - \frac{2}{3}\right)}} 
     - \frac{2}{3} \frac{\Omega_{2}}{\Omega t^2} \right] \delta_{2} - 
     \frac{2}{3} \, \frac{\Omega_{1}}{\Omega} \frac{\delta_{1}}{t^2} = 0 \;.
\end{array}
\right.
\end{equation}
For an arbitrary $\gamma_{i}$ $\left( \geq \frac{4}{3} \, \right.$for 
physical reasons $\left. \cite{wein} \right)$ , system (8) may be 
numerically studied ; we give and 
discuss a set of exact analytic solutions concerning the foolowing situations 
of some physical intereset : $\gamma_{1} = \gamma_{2} > \frac{4}{3} \, , \, 
v^2_{1} = v^2_{2} \;$ ; $\gamma_{1} = \gamma_{2} = \frac{4}{3} \, , \, 
v^2_{1} \neq v^2_{2}$.\\
For solutions see the journal-ref.

\end{document}